\documentstyle[amssymb,aps,12pt]{revtex}

\begin{document}
\title{DISTRIBUTION OF CARBON ATOMS IN IRON-CARBON {\it fcc} PHASE: AN EXPERIMENTAL
AND THEORETICAL STUDY}
\author{K. F. Laneri, J. Desimoni}
\address{Departamento de F\'{\i }sica, Facultad de Ciencias. Exactas, UNLP,\\
IFLP-CONICET C.C. 67, 1900 La Plata, Argentina}
\author{G. J. Zarragoicoechea}
\address{Instituto de F\'{\i }sica de L\'{\i }quidos y Sistemas Biol\'{o}gicos\\
(CICPBA-UNLP) 59 N$^o$ 789, C.C. 565, 1900 La Plata, Argentina}
\author{A. Fern\'{a}ndez-Guillermet}
\address{Consejo Nacional de Investigaciones Cient\'{\i }ficas y T\'{e}cnicas, Centro%
\\
At\'{o}mico Bariloche, 8400 Bariloche, Argentina}
\maketitle

\begin{abstract}
This paper presents an experimental and theoretical study of the
distribution of carbon atoms in the octahedral interstitial sites of the
face-centered cubic ({\it fcc}) phase of the iron-carbon system. The
experimental part of the work consists of M\"{o}ssbauer measurements in Fe-C
alloys with up to about 12 atomic percent C, which are interpreted in terms
of two alternative models for the distribution of C atoms in the
interstitial sites. The theoretical part combines an analysis of the
chemical potential of C based on the quasichemical approximation to the
statistical mechanics of interstitial solutions, with three-dimensional
Monte Carlo simulations. The latter were performed by assuming a gas like
mixture of C atoms and vacancies (Va) in the octahedral interstitial sites.
The number of C-C, C-Va and Va-Va pairs calculated using Monte Carlo
simulations are compared with those given by the quasichemical model.
Furthermore, the relative fraction of the various Fe environments were
calculated and compared with those extracted from the M\"{o}ssbauer spectra.
The simulations reproduce remarkably well the relative fractions obtained
assuming the Fe$_8$C$_{1-y}$ model for M\"{o}ssbauer spectra, which includes
some blocking of the nearest neighbour interstitial sites by a C atom. With
the new experimental and theoretical information obtained in the present
study, a critical discussion is reported of the extent to which such
blocking effect is accounted for in current thermodynamic models of the Fe-C 
{\it fcc} phase.

\end{abstract}

\section{INTRODUCTION}

The physical properties of the austenite solid solution phase have been
studied extensively over the years in connection with, e.g., the assessment
and understanding of the phase diagram [1-6], the diffusion controlled [7]
and martensitic phase transitions in Fe-C alloys [8]. In austenite the iron
atoms are arranged in a close-packed face-centred cubic ({\it fcc}) lattice,
and the C atoms occupy a limited number of the octahedral interstices which
are located at the centres and at the mid-points of the edges of the unit
cubes, these two positions being crystallographically equivalent (Fig.1)
[9]. Accordingly, various models of austenite have been proposed which are
based on assuming two sublattices, one for the Fe atoms and the second one
for the mixture of carbon atoms (C) and vacant octahedral interstices (Va).
The general theme of the present paper is the distribution of the C atoms in
the interstitial sites, as revealed by three complementary sources of
information, viz., thermodynamic properties, M\"{o}ssbauer experiments and
Monte Carlo simulations.

In the ideal solution model (ISM) for thermodynamic properties of austenite
it is assumed that C and Va distribute themselves at random in the
octahedral interstitial sites, the amount of which is equal to the number of
Fe atoms. On these basis, the thermodynamic activity ($a_c$) of C in an
ideal mixture of N$_C$ carbon atoms with N$_{Fe}$ iron atoms is shown to be
proportional to the ratio $y_c/(1-y_c)$, where $y_c$ = N$_C$/N$_{Fe}$, and $%
1-y_c$ represent the fraction of occupied and of empty interstitial sites,
respectively [10-13]. Since the experimental $a_c$ in austenite deviates
positively from the ISM, many approaches have been proposed to account for
the non-ideal behaviour [14-22]. The reader is referred to ref. [23] for a
recent review of the work of most relevance for the present study. In the
strict version of the approach known as the hard-blocking excluded-sites
model (HBESM) it is assumed that the presence of a solute atom blocks the
occupancy of a certain number ($b$) of the nearest neighbor interstitial
sites (NNIS), so that a site is either blocked or is available for the
mixing of C atoms and Va [9, 15 , 23]. Further, if the mixing in the
non-blocked sites occurs at random, the $a_c$ in austenite becomes
proportional to $y_c/[1-(1+b)y_c]$ [23]. Frequently, $b$ has been treated as
an adjustable parameter, identified with the value to be inserted in the
expression for $a_c$ in order to reproduce the experimental data
[9,19,22,23]. Alternatively, some theoretical studies have been reported
which suggest that $b$ should in fact be treated as composition dependent
[18,19]. A different approach will be explored in the present work, which is
based on combining two theoretical methods. First, we will adopt the
quasi-chemical approximation (QCA) to the statistical mechanics of
interstitial solutions [12,24-27]. In the QCA all interstitial sites are
available for mixing, but the C atoms are regarded as exerting a repulsive
force on each other, so that they enter adjacent interstitial positions less
frequently as would be the case if their distribution were random. Thus the
QCA will allow us to treat soft-blocking effects in austenite. The key
parameters in this treatment are the energies of formation of the C-C and
C-Va pairs, which will be accurately determined by analysing $a_c$ data.
Secondly, we will perform Monte Carlo (MC) simulations for various values of
the N$_C$/N$_{Fe}$ ratio, using the pair formation energies from the
quasi-chemical analysis. In this way, the average distribution of
interstitials around a given C atom will be studied as function of
composition. In particular, the average number of empty NNIS will be
determined for various alloys, and compared with the results of previous
studies, as well as with information extracted from M\"{o}ssbauer
experiments.

In the analysis of the M\"{o}ssbauer spectra of austenite various
assumptions about the distribution of C in the octahedral interstitial sites
have been proposed [28,29]. In particular, a model has been suggested for
dilute solutions in which the 12 NNIS of a C atom are excluded [28]. In this
case the C atoms occupy only the centre of the cubes of a structure with the
formula Fe$_8$C$_{1-y}$, so that three possible environments for Fe atoms,
associated to different hyperfine interaction may be distinguished, which
are shown schematically in Fig.2, viz.,

a. Fe atoms without nearest neighbor and next nearest neighbor C atoms,
associated to the singlet $\Gamma _{00}$.

b. Fe atoms without nearest neighbour but with $n$ next nearest neighbour C
atoms ($n=1-4$), ascribed to the singlet $\Gamma _{0n}$.

c. Fe atoms with one C atom nearest neighbour but without next nearest C
neighbours, related to doublet $\Gamma _{10}$.

Alternatively, a model [29] has been proposed in which all the octahedral
sites of the {\it fcc} structure are available for occupation and no
assumptions are made on the distribution of the C atoms in the second
interstitial shell. The various Fe-C environments involved in this type of
model are shown schematically in Fig.3, viz.,

a. Fe atoms without nearest neighbours C atoms, associated to the singlet $%
\Gamma _0$.

b. Fe atoms with one nearest neighbor C atom, or Fe atoms with two C atoms
nearest neighbours at 90$^0$ from each other, related to the doublet $\Gamma
_1$.

c. Fe atoms with two C atoms placed at opposite nearest sites (180$^0$),
ascribed to the doublet $\Gamma _2$.

The purpose of the present paper is to provide new experimental and
theoretical information on the distribution of C atoms in the octahedral
sites of the austenite phase. The work proceeds as follows. First,
M\"{o}ssbauer experiments are performed on a series of alloys with up to 12
at.\%.

\section{EXPERIMENTAL}

\subsection{Alloys, samples and heat-treatments}

Five Fe-C compacted graphite alloys were prepared in a medium frequency
induction furnace using the sandwich technique in ladle to treat the liquid
metal. The necessary amount of Si, Mg, Ce and Ca was added to obtain
compacted graphite morphology. The final alloy contained 3.40 wt.\%C, 2.35
wt.\%Si, 0.58 wt.\%Mn, 0.04 wt.\%Cu, 0.01 wt.\%P, and 0.02 wt.\%S.

Samples of 20 mm diameter and 3 mm thickness were taken from ''Y-shape''
blocks (ASTM A-395) cast in sand moulds.Samples S1, S2, S3, S4 and S5 (see
below) were annealed at 1173K during 30 minutes, quenched in a salt bath and
held at 623K during 1, 3, 4, 5 and 10 minutes respectively. The samples for
M\"{o}ssbauer spectroscopy were prepared by conventional grinding techniques
to reduce their thickness down to about 70mm, using diamond paste of 6, 1
and 0.1mm for final polishing.

\subsection{X-ray measurements}

X-ray measurements were performed in a Philips PW1710 diffractometer using
the monochromatic {\it K}$_\alpha $ radiation of Cu, in Brag Brentano's
geometry, with a step mode collection of 0.02, 10s by step, with 2$\theta $
ranging from 39$^0$ to 98$^0$. The X-ray patterns, presented in Fig.4, were
analysed with the Rietveld method [30]. The actual C concentration in the
samples was determined by combining the lattice-parameters ($a$) extracted
from the diffraction patterns with the known a versus composition relation
for {\it fcc} Fe-C alloys [28]. The resulting lattice-parameter and the
inferred $y_c$ values for the various alloys are listed in Table I.

\subsection{M\"{o}ssbauer experiments}

M\"{o}ssbauer spectra were taken in a transmission geometry using a 57Co%
\underline{{\it Rh}} source of approximately 5mCi intensity and recorded in
a standard 512 channels conventional constant acceleration spectrometer. In
order to analyse in detail the austenite pattern, the spectra were taken in
the velocity range between - 2 and + 2 mm/s. Velocity calibration was
performed against a 12 mm thick $\alpha $ - Fe foil. All isomer shifts were
referred to this standard at 298K. The spectra were fitted to Lorentzian
line shapes using a non-linear least-squares program with constraints. For
the effective thickness of the samples analysed no Voigt line-shape
correction was necessary [31].

The results of the M\"{o}ssbauer experiments are shown in Fig.5. The central
subspectra were associated to austenite, whereas the external lines on the
spectra to ferrite/martensite phases [28]. The hyperfine parameters and the
relative fractions f$_{lm}$ ($l,m$ = number of C atoms in the first and
second coordination shell, respectively) associated to the various Fe
environments obtained using the models [28,29] referred to in Sect.I are
listed in Table II. Concerning the second model [29], the contribution to
the spectra of the doublet $\Gamma _2$ associated to Fe sites with two C
atoms placed in opposite interstitial sites resulted undetectable by the
present technique.

\section{THEORETICAL}

\subsection{Monte Carlo simulations}

The austenite interstitial solid solution is described as a lattice gas of N$%
_C$ carbon atoms and N$_{Va}$ vacancies, distributed in the N$_C$+ N$_{Va}$
= N=N$_{Fe}$ octahedral interstitial sites of the {\it fcc} structure (Fig.
1) associated to N$_{Fe}$ iron atoms. The occupancy of the first and second
interstitial shell was accounted for in three-dimensional MC simulations to
calculate the number $n_{ij}$ ($i,j$ = C or Va) of C-C, C-Va and Va-Va
pairs, the relative fractions f$_{lm}$ associated to the different Fe
environments, and the number C$_{i0}$ of C atoms having $i$ C atoms in the
first interstitial coordination shell and none in the next interstitial
shell.

A Fortran 77 routine using the Monte Carlo method, an Ising-type Hamiltonian
and periodic boundary conditions was developed. Metropolis method was used
to define the probability of the C jumps. A randomly chosen C atom has the
probability $P$ to jump to an empty interstitial neighbouring site, also
randomly chosen, viz.,

\begin{equation}
P= 
{\exp [(\varepsilon _{Ti}-\varepsilon _{Tf})/RT]\text{ \ if }\varepsilon _{_{Tf}}\text{\mbox{$>$} }\varepsilon _{_{Ti}} \atopwithdelims\{\} 1\text{ \qquad \qquad \qquad \qquad \ if }\varepsilon _{_{Tf}}\leq \varepsilon _{_{Ti}}}%
\eqnum{1}
\end{equation}
\newline
where $\varepsilon _{Ti}$ and $\varepsilon _{Tf}$, are the initial and final
total energies, respectively, calculated using the relation $\varepsilon
_T=n_{C-C}\Delta \varepsilon $. Here $\Delta \varepsilon $ = 2 $\varepsilon
_{C-Va}$ - $\varepsilon _{C-C}$ is the energy of formation of a C-C pair of
nearest neighbour C atoms relative to the individual C atoms (see below), $%
n_{C-C}$ is the number of C-C pairs, $\varepsilon _{C-C}$ and $\varepsilon
_{C-Va}$ are the interaction energies of the C-C and C-Va pairs,
respectively, $R$ is the gas constant and $T$ is the temperature in Kelvin.
If the atom movement decreases the total energy, the jump is allowed ($P$%
=1), but if the total energy increases, the jump is allowed with a
probability $P=exp[(\varepsilon _{Ti}-\varepsilon _{Tf})/RT]$. The Fe atoms
remain still during the simulation, and their positions were only used to
calculate the number of $n_{ij}$ pairs and the relative fractions f$_{lm}$
associated to the different Fe environments.

In order to study the convergence of the results, cells of 4$^3$, 6$^3$, 8$%
^3 $ and 10$^3$ were used. For simulations using cell sizes of 6$^3$ and
higher the $n_{ij}$ and f$_{lm}$ fractions did not vary, hence cells of 864
Fe atoms and the corresponding number of C atoms were employed to decrease
the calculation time. For all C concentrations, the equilibrium of the
system was attained approximately at three MC steps, where a MC step is
defined as N$_C$ attempts of movement of a C atom.

Finally, the occupation of the interstitial sites was characterised using
the average number $z$ of empty NNIS, which was calculated from the MC
results as follows:

\begin{equation}
z=\frac{\sum\limits_{i=0}^{12}C_{i0}(12-i)}{\sum\limits_{i=0}^{12}C_{i0}} 
\eqnum{2}
\end{equation}

\subsection{Quasichemical model calculations}

The energy of formation of a C-C pair that enters in the MC calculation was
determined by analysing experimental $a_c$ data in terms of the QCA to the
statistical mechanics of the Fe-C solutions developed by Bhadeshia [27].
This formalism yields for the activity $a_c$ 
\begin{equation}
a_c=\frac{y_c}{1-y_c}\exp [\frac{\Delta \text{G}_c}{RT}]\left\{ \left( \frac{%
y_c}{1-y_c}\right) ^2\left( \frac{1-y_c-\frac{\overline{\lambda }}{N_{Va}}}{%
1-y_c}\right) \right\} ^{\frac{-Z}2}\exp [\frac{-Z\Delta \varepsilon }{2RT}]
\eqnum{3}
\end{equation}
where $Z$ (=12) is the number of NNIS, and $\Delta $G$_c$ is the Gibbs
energy of C in austenite relative to graphite. The value of the parameter $%
\overline{\lambda }$ that minimise the Gibbs energy is: 
\[
\overline{\lambda }=\frac{N_{Va}}{2\sigma }\left\{ 1-\left[ 1-4\sigma
y_c\left( 1-y_c\right) ^{\frac 12}\right] \right\} 
\]
with 
\[
\sigma =1-\exp \left[ \frac{-\Delta \varepsilon }{RT}\right] 
\]

A linear approximation of Eq.3, appropriate for describing the dilute
solution range was fitted to carbon activity data measured at 1423K [22]. A
least-squares fit of the equation: 
\[
RT\ln \left[ a_c\frac{1-y_c}{y_c}\right] =y_cZ\Delta \varepsilon +\Delta 
\text{G}_c 
\]
which is shown in Fig.6 yielded $\Delta $G$_c$=4451$_{25}$ cal/mol, and $%
\Delta \varepsilon $=1492$_{39}$ cal/mol. This one was adopted in the MC
simulations.

The number of pairs $n_{ij}$ calculated for $0<y_c<1$ using the
quasichemical formalism (Table III) are plotted in Fig. 7 using lines. For
comparison the $n_{ij}$ determined in the MC calculations using $\Delta
\varepsilon $ = 1492 cal/mol are plotted using symbols. The inset there
gives a comparison for the composition range corresponding to the
experimental solubility of C in austenite, viz., $y_c<0.1$. There is a very
good agreement between the QCA and the MC predictions for $n_{ij}$, which
encourages a discussion of the MC results for the relative fractions f$_{lm}$
of the various Fe environments, as functions of $y_c$.

\section{DISCUSSION}

\subsection{Monte Carlo versus M\"{o}ssbauer results}

The MC results for the f$_{lm}$ as functions of $y_c$ are plotted in Fig. 8.
According to the present simulations the main contributions to the
M\"{o}ssbauer spectra originate in the Fe environments without C atoms in
the first interstitial shell (f$_{00}$ and f$_{0n}$). Next in importance is
the contribution of environments with one C atom in the first interstitial
shell and none in the second (f$_{10}$). Further, the simulations show that
the contribution of Fe atoms having the first and second interstitial shells
occupied (f$_{nm}$) is not negligible, which indicates that this kind of Fe
environments should be accounted for in fitting M\"{o}ssbauer spectra.
Figure 8 also demonstrates that the relative fraction associated to Fe atoms
with more than one C atom in the first interstitial shell and without C
atoms in the second shell (f$_{n0}$) is negligible. This result contradicts
the assumption of one of the models [29] developed to interpret the
M\"{o}ssbauer pattern of the austenite, which was reviewed in Sect.I. In
fact, we find that the contribution of Fe environments with two C atoms
either at 90$^0$ or 180$^0$ is negligible.

In order to compare the MC results with the results of analysing the present
M\"{o}ssbauer spectra (Table II), the relative fractions f$_{00}$ and the
sum f$_0$ = f$_{00}$ + f$_{0n}$ were chosen as the key quantities in the Fe$%
_8$C$_{1-y}$ model [28] and the random model [29], respectively. The reason
for this choice is that Fe environment with C atoms in the first and the
second interstitial shells have not been considered by any of the models
proposed to reproduce the M\"{o}ssbauer spectra [28,29]. Moreover, we have
also shown that the fraction of Fe environments with two C atoms in the
first shell is negligible according to MC results. Hence, the only fractions
that can be determined without ambiguity for these models are f$_{00}$ and f$%
_0$, respectively.

A comparison between the f$_{00}$ and f$_0$ versus composition values
extracted from the M\"{o}ssbauer spectra (open symbols) and the MC
simulations (filled symbols) is presented in Fig. 9. According to Fig. 9 the
''random model'' underestimates significantly the contribution of the Fe
environments without nearest neighbour C atoms. This suggests that in this
composition range, some blocking effect of the NNIS in austenite should be
accounted for. In agreement with this expectation, the MC results fit
remarkably well with the f$_{00}$ values extracted from M\"{o}ssbauer
experiments when the Fe$_8$C$_{1-y}$ model is adopted, i.e., the model
including some blocking effect of the interstitial sites.

Finally, MC simulations have been reported previously [32] which are based
on extracting from M\"{o}ssbauer measurements a weak C-C repulsion in the
first coordination shell (w$_1$ = 830 cal/mol) and a stronger one (w$_2$ =
1730 cal/mol) in the second shell. Such calculations were interpreted
[29,32] as indications that the Fe$_8$C$_{1-y}$ model is not adequate to
represent the austenite phase. Similar results (w$_1$ = 830 cal/mol) were
arrived at in ref. [29] by assuming that pairs of C atoms can occupy the
first interstitial shell either at 90$^0$ or at 180$^0$. The present MC
results, based on energies extracted from thermodynamic data, do not support
these results.

\subsection{Account of blocking effects}

In Fig. 10 the average number of empty NNIS $z$ obtained from the MC
simulations is plotted as function of $y_c$ (symbols). The $z$ values
corresponding to the composition of the present experimental alloys are
plotted using empty symbols. The dashed line in this graphic refers to the $%
z $ value corresponding to a random mixture, viz., $z=12(1-y_c)$. The empty
symbols in Fig.10 indicate that already in alloys with $y_c=0.05$ the NNIS
of the C atoms are, on the average, less occupied than in a random mixture.
This fact is in qualitative agreement with the ideas behind the
excluded-sites model, which motivates the following analysis of the blocking
effects in models for $a_c$ in austenite.

It has recently been pointed out [23] that in a strict hard-blocking model,
empty sites must be interpreted as blocked sites. This implies that the $b$
parameter of the HBESM (Sect.I) should be considered as equal to $z$
(Fig.10). Two consequences of such interpretation will be discussed. The
first consequence is that the MC results in Fig.10 cannot be represented
using the HBESM unless the $b$ parameter is allowed to vary with
composition. In qualitative agreement with this, Oates et al. [21]
interpreted their own $z$ values from MC calculations as a composition
dependent $b$ parameter, decreasing with the increase in the C content.
However, the excluded sites model does not explain the composition
dependence of $b$, which has stimulated some attempts to improve the simple
picture by invoking, e.g., overlapping of the sites excluded by different
interstitial atoms [17,19,21,23]. The second consequence of the current [23]
interpretation of the hard-blocking is that Fig. 10 yields $b$ (=$z$) 
\mbox{$>$}%
10 in the composition range $y_c<0.2$. However, these values cannot be
reconciled with those extracted from experimental $a_c$ data. The latter are
integral and non-integral $b$ values falling in the range 3 
\mbox{$<$}%
$b$ 
\mbox{$<$}%
5 [9,19,22,23].

In view of these facts we conclude that the strict form of the
excluded-sites model (Sect.I) does not seem able to account for the present
blocking effects by using the same $b$ values which are known to reproduce
the experimental $a_c$ data. It is also evident that a more realistic
account of such effects would require abandoning the one-parameter formula
for $a_c$, i.e., what has been considered as the main advantage of the HBESM
[23].

\section{SUMMARY AND CONCLUDING REMARKS}

In the present study the energy parameters describing the interstitial
solution of C in the {\it fcc} phase of Fe have been obtained by analyzing
experimental thermodynamic data in terms of the quasichemical approximation.
These parameters have been used as input information in Monte Carlo
simulations, and various key quantities have been obtained. In particular,
the composition dependence of various f$_{ij}$ ratios, describing the
relative weight of the various Fe configurations contributing to the
M\"{o}ssbauer spectra of Fe-C austenite, have been predicted and compared
with those derived by modelling the M\"{o}ssbauer spectra. In this way, two
alternative models [28,29] for the contribution of the Fe environments have
been tested. The present comparison between M\"{o}ssbauer and theoretical
results indicates that a description similar to the Fe$_8$C structure [28]
should be preferred for the Fe-C austenite phase. Such a model [28] is
usually associated to the blocking of some of the nearest neighbour
interstitial sites by a C atom. However, the present results cannot be
accounted for by the simplest hard-blocking excluded sites model, often used
to provide a one-parameter formula for the activity of C in austenite. We
believe that the picture of blocking effects in austenite emerging from the
present study should be useful in further attempts to refine the current
models for thermodynamics of interstitial solutions.

{\LARGE ACKNOWLEDGMENTS}

This work was partially supported by Consejo Nacional de Investigaciones
Cient\'{\i }ficas y T\'{e}cnicas (CONICET), PICT 1277, PICT 034517 and
PICT-99-03-6507 of the Agencia Nacional de Promoci\'{o}n Cient\'{\i }fica y
Tecnol\'{o}gica (ANPCyT). G. J. Z. is member of Carrera del investigador
CICPBA.\newpage

{\LARGE REFERENCES}

[1] L. S. Darken and R. W. Gurry, ''Physical Chemistry of Metals'', McGraw
Hill, New York, 1953.

[2] M. Benz and J. Elliott, Trans. AIME {\bf 221}, 323 (1961).

[3] H. Harvig, Jernkontor.Ann. {\bf 155}, 157 (1971).

[4] J. Chipman, Met. Trans. {\bf 3}, 55 (1972) .

[5] J. Agren, Met. Trans. {\bf 10A}, 1847 (1979).

[6] P. Gustafson, Scand.Journ.Met. {\bf 14}, 259 (1985).

[7] ''Decomposition of Austenite by Diffusional Processes'', V. F. Zackay
and H. I. Aaronson, editors. Interscience Publishers, New York, 1962.

[8] ''Martensite'', G. B. Olson and W. S. Owen editors. American Society for
Metals International, 1992.

[9] L. Kaufman, S. V. Radcliffe and M. Cohen in ref[7], p.313.

[10] J. R. Lacher, Proceed. Royal Society (London) Ser.A {\bf 161}, 525
(1937) .

[11] J. R. Lacher, Proceed. Camb. Phil. Soc. {\bf 33}, 518 (1937).

[12] R. Fowler and E.A.Guggenheim, ''Statistical Thermodynamics'', Cambridge
Univ. Press, Cambridge, 1956.

[13] R. W. Gurney, ''Introduction to Statistical Thermodynamics'', McGraw
Hill, New York, 1949.

[14] R. B. McLellan, in ''Phase Stability in Metals and Alloys'', P. S.
Rundman, J. Stringer and R. I. Jaffee, editors. McGraw Hill Co, New York,
1968.

[15] R. Speiser and J. W. Spretnak, Trans. ASM {\bf 47}, 493 (1955).

[16] K. A. Moon, Trans.AIME {\bf 227}, 1116 (1963) .

[17] R. B .McLellan, T. L. Garrard, S. J. Horowitz and J. A. Sprague, Trans
AIME {\bf 239}, 528 (1967).

[18] P. T. Gallagher, J. A. Lambert and W. A. Oates, Trans.AIME {\bf 245},
887(1969).

[19] H. M. Lee, Metall.Trans. {\bf 5}, 787 (1974).

[20] M. Hillert and L-I. Staffansson, Acta Chem.Scand. {\bf 24}, 3618 (1970).

[21] W. A. Oates and T. B. Flanagan, Journ.Mat.Science {\bf 16}, 3235 (1981).

[22] S. Ban-ya, J. F. Elliott, and J. Chipman, Trans. AIME {\bf 245}, 1199
(1969).

[23] M. Hillert, Zeits. Metallkde, {\bf 90}, 60 (1999).

[24] E. A. Guggenheim, ''Mixtures'',Oxford Univeristy Press, 1952.

[25] L. S. Darken and R. P. Smith, Journ.Amer.Chem.Soc. {\bf 68}, 1172
(1946).

[26] R. B. Mc Lellan and W. W. Dunn, J. Phys. Chem. Solids {\bf 30}, 2631
(1969).

[27] H.K.D.H. Badheshia, Mater.Sci. and Technology {\bf 14}, 273 (1998).

[28] O.N.C.Uwakweh, J.P.Bauer and J.M.G\'{e}nin, Metall. Trans. {\bf A21}
589 (1990).

[29] K. Oda, H. Fujimura, H. Ino, J. Phys.:Condens Matter {\bf 6}, 679
(1994).

[30] R. A. Young, in: The Rietveld method, (International Union of
Crystallography, Oxford, University Press, 1993).

[31] D. G. Rancourt, A. M. Mc Donald, A. E. Lalonde and J. Y. Ping: American
Mineralogist {\bf 78}, 1(1993).

[32] A. L. Sozinov, A.G. Balanyuk and V.G. Gavriljuk, Acta Mater. {\bf 45},
225 (1997).\newpage

{\LARGE FIGURE CAPTIONS}

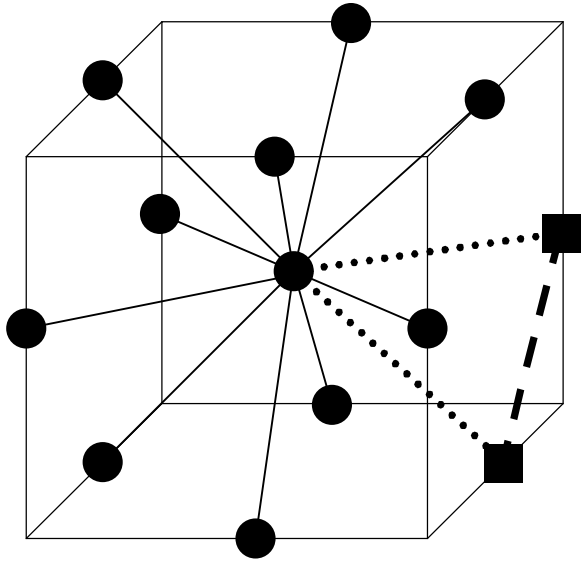
Figure 1: Interstitials sites of the austenite {\it fcc} phase. a) $%
\blacksquare $ : C atom, b) $\bullet $ : Vacancy, c) dotted line: C-Va pair,
d) solid line: Va-Va pair and e) dashed line: C-C pair. Fe atoms are placed
in the corners of the cube and in the face centres.

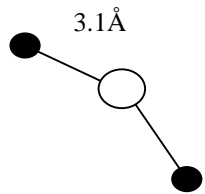
Figure 2: Fe environments in the Fe$_8$C$_{1-y}$ model [28]. Filled and open
circles correspond to C and Fe atoms, respectively.

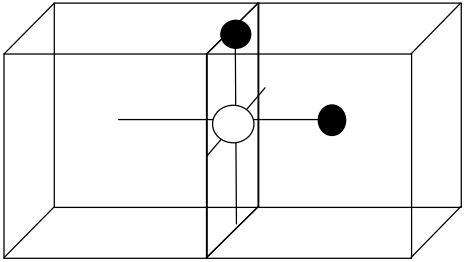
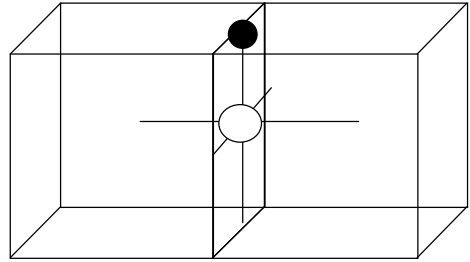
Figure 3: Fe environments in the random model [29]. Filled and open circles
correspond to C and Fe atoms, respectively.

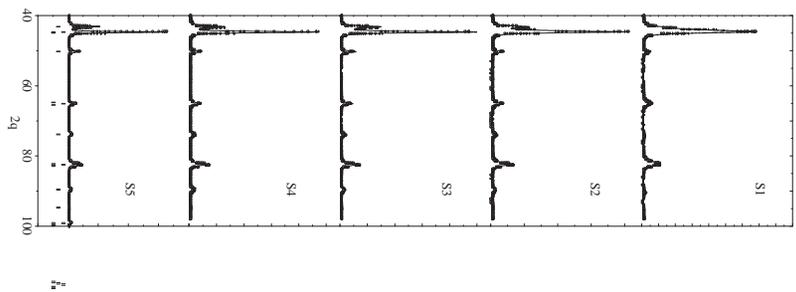
Figure 4: X-Ray diffractograms of samples S1 to S5. The bottom bar diagrams
indicate from top to bottom: ferrite, austenite, martensite and C graphite.

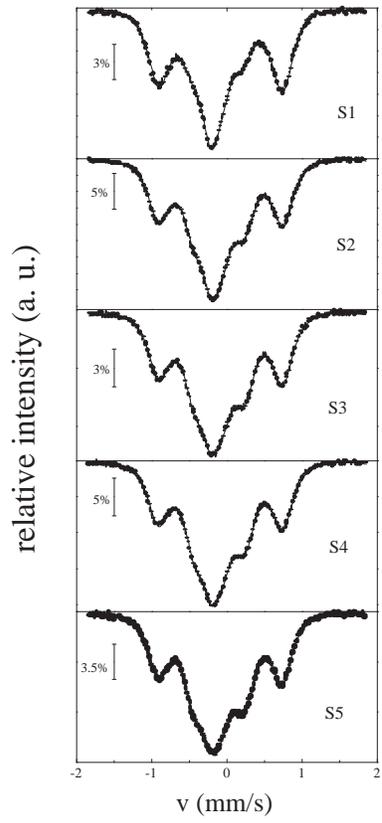
Figure 5: M\"{o}ssbauer spectra recorded for samples S1 to S5.

Figure 6: Linear fit for activity data of ref.[22]. The values $\Delta
\varepsilon $ = 1492$_{39}$ cal/mol and $\Delta $G$_c$ = 4451$_{25}$ cal/mol
were determined.

Figure 7: The number n$_{ij}$ (i, j = C or Va) of pairs C-C, C-Va and Va-Va.
Squares, triangles and circles represent Va-Va, C-C and C-Va pairs
respectively, obtained from Monte Carlo simulations using $\Delta
\varepsilon $=1492 cal/mol. a) dash-dotted line: Va-Va, b) dashed line: C-C
pairs and c) dotted line: C-Va pairs, calculated using the quasichemical
model with the same $\Delta \varepsilon $ value. The inset gives a
comparison for the composition range corresponding to the experimental
solubility of C in austenite, viz., y$_c$ 
\mbox{$<$}%
0.1.

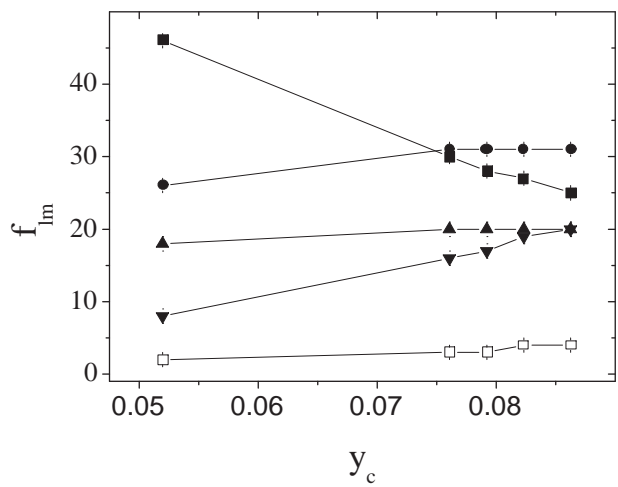
Figure 8: The relative fractions f$_{lm}$ (l, m = number of C atoms in the
first and second coordination shell, respectively) associated to the various
Fe environments obtained using Monte Carlo simulations, as functions of C
content.

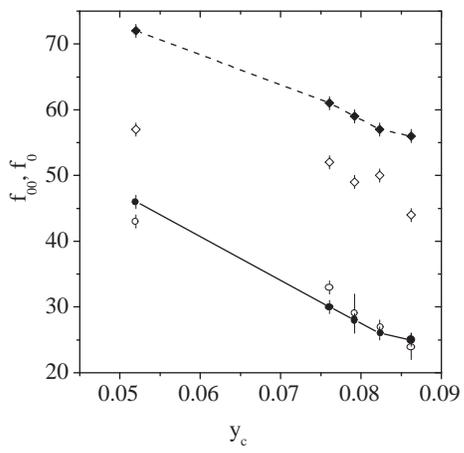
Figure 9: The relative fractions f$_{00}$ (circles) and f$_0$ (diamonds)
associated to the various Fe environments obtained from M\"{o}ssbauer data
(open symbols) using the models [28,29] referred to in Sect.I, compared with
results from Monte Carlo simulations (filled symbols).

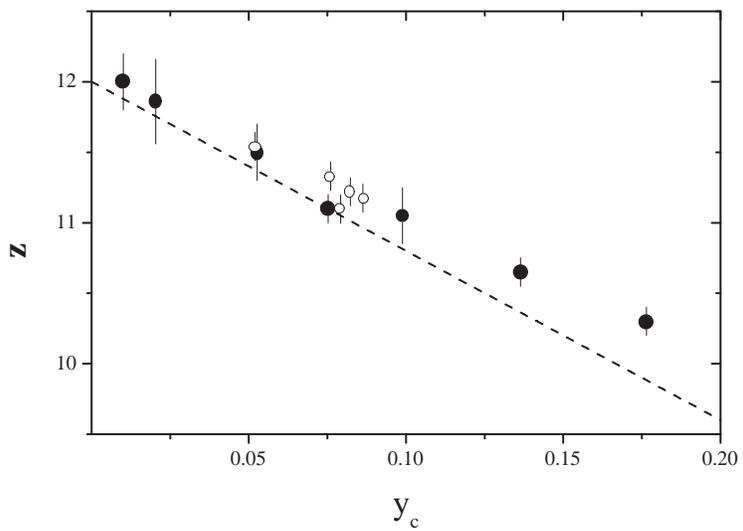
Figure 10: The average number $z$ of empty nearest neighbour interstitial
sites of a C atom in austenite calculated as a function of composition using
Monte Carlo simulations.

\smallskip

{\LARGE TABLE CAPTIONS}

Table I: Lattice parameter ({\it a}) determined from the diffractograms
using Rietveld [30] and corresponding to the samples S1, S2, S3, S4 and S5.
The C content was determined using the empirical relation of ref.28.

Table II: Hyperfine parameters and relative fractions of the different Fe
environments found in austenite using the models of refs. 28 and 29.

Table III: Pair interaction energy and number of n$_{ij}$ pairs in the
austenite phase obtained using the quasichemical model [27].\newpage
\newpage Table I

\begin{tabular}{|l|l|l|}
\hline
Sample & Cell constant {\it a} (\AA ) & $y_c$ \\ \hline
S1 & 3.610$_1$ & 0.052$_1$ \\ \hline
S2 & 3.626$_1$ & 0.076$_1$ \\ \hline
S3 & 3.628$_1$ & 0.079$_1$ \\ \hline
S4 & 3.630$_1$ & 0.082$_1$ \\ \hline
S5 & 3.632$_1$ & 0.086$_1$ \\ \hline
\end{tabular}

\newpage

Table II

\begin{tabular}{|c||c|c|c|c|c|c|c||c|c|c|c|c|}
\cline{2-13}
\multicolumn{1}{c||}{} & \multicolumn{2}{||c|}{$\Gamma _{00}$} & 
\multicolumn{2}{|c|}{$\Gamma _{0n}$} & \multicolumn{3}{|c||}{$\Gamma _{10}$}
& \multicolumn{3}{||c}{$\Gamma _1$} & \multicolumn{2}{|c|}{$\Gamma _0$} \\ 
\hline
Sample & $%
{\delta  \atop mm/s}%
$ & $%
{f_{00} \atop \%}%
$ & $%
{\delta  \atop mm/s}%
$ & $%
{f_{0n} \atop \%}%
$ & $%
{\Delta  \atop mm/s}%
$ & $%
{\delta  \atop mm/s}%
$ & $%
{f_{10} \atop \%}%
$ & $%
{\Delta  \atop mm/s}%
$ & $%
{\delta  \atop mm/s}%
$ & $%
{f_1 \atop \%}%
$ & $%
{\delta  \atop mm/s}%
$ & $%
{f_0 \atop \%}%
$ \\ \hline
S1 & -0.1 & 43$_1$ & 0.05$_1$ & 16$_1$ & 0.66$_1$ & -0.01$_1$ & 41$_2$ & 0.61%
$_1$ & -0.01$_1$ & 43$_2$ & -0.07$_1$ & 57$_1$ \\ \hline
S2 & -0.1 & 33$_1$ & 0.05$_1$ & 23$_1$ & 0.67$_1$ & 0.01$_1$ & 44$_2$ & 0.62$%
_1$ & 0.01$_1$ & 48$_1$ & -0.05$_1$ & 52$_1$ \\ \hline
S3 & -0.1 & 29$_7$ & 0.06$_1$ & 23$_1$ & 0.67$_1$ & 0.01$_1$ & 48$_3$ & 0.63$%
_1$ & 0.02$_1$ & 51$_1$ & -0.04$_1$ & 49$_1$ \\ \hline
S4 & -0.1 & 27$_1$ & 0.06$_1$ & 21$_1$ & 0.67$_1$ & 0.01$_1$ & 52$_2$ & 0.63$%
_1$ & 0.02$_1$ & 50$_1$ & -0.04$_1$ & 50$_1$ \\ \hline
S5 & -0.1 & 24$_2$ & 0.05$_1$ & 25$_2$ & 0.67$_1$ & 0.01$_1$ & 51$_3$ & 0.63$%
_1$ & 0.02$_1$ & 56$_1$ & -0.03$_1$ & 44$_1$ \\ \hline
\end{tabular}
\newline

\smallskip

\newpage \smallskip

Table III

\begin{tabular}{|l|l|l|}
\hline
Kind of pair & Number of pairs ( n$_{ij}$) & Energy per pair \\ \hline
Va-Va & n$_{Va-Va}$ = $\frac 12$ Z N (1-y$_c$-$\lambda $) & 0 \\ \hline
C-Va+Va-C & n$_{C-Va}$ = Z N $\lambda $ & $\varepsilon _{C-Va}$ \\ \hline
C-C & n$_{C-C}$ = $\frac 12$ Z N (y$_c$-$\lambda $) & $\varepsilon _{C-C}$
\\ \hline
\end{tabular}

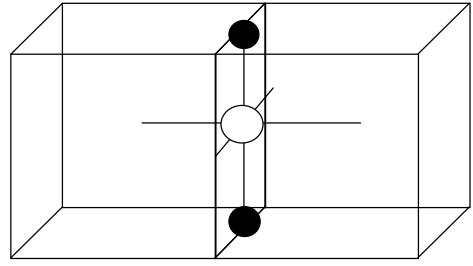
\end{document}